\documentclass[traditabstract]{aa}  
\usepackage{graphicx}
\usepackage{txfonts}

\begin{document}

\title{A progenitor binary and an ejected mass donor remnant of faint type Ia supernovae}

\author{S. Geier \inst{1,2}
   \and T. R. Marsh \inst{3}
   \and B. Wang \inst{4}
   \and B. Dunlap \inst{5}
   \and B. N. Barlow \inst{6}
   \and V. Schaffenroth \inst{2,7}
   \and X. Chen \inst{4}
   \and A. Irrgang \inst{2}
   \and P. F. L. Maxted \inst{8}
   \and E. Ziegerer \inst{2}
   \and T. Kupfer \inst{9}
   \and B. Miszalski \inst{10,11}
   \and U. Heber \inst{2}
   \and Z. Han \inst{4}
   \and A. Shporer \inst{12,13,14}
   \and J. H. Telting \inst{15}
   \and B. T. G\"ansicke \inst{3}
   \and R. H. \O stensen \inst{16}
   \and S. J. O'Toole \inst{17}
   \and R. Napiwotzki \inst{18} 
   }

\offprints{S.\,Geier,\\ \email{sgeier@eso.org}}

\institute{European Southern Observatory, Karl-Schwarzschild-Str.~2, 85748 Garching, Germany
\and Dr. Karl Remeis-Observatory \& ECAP, Astronomical Institute, Friedrich-Alexander University Erlangen-N\"urnberg, Sternwartstr. 7, D 96049 Bamberg, Germany
\and Department of Physics, University of Warwick, Conventry CV4 7AL, United Kingdom 
\and Key Laboratory of the Structure and Evolution of Celestial Objects, Yunnan Observatory, Chinese Academy of Sciences, Kunming 650011, China
\and Department of Physics and Astronomy, University of North Carolina, Chapel Hill, NC 27599-3255, USA
\and The Pennsylvania State University, 525 Davey Lab, University Park, PA 16802, USA 
\and Institute for Astro- and Particle Physics, University of Innsbruck, Technikerstr. 25/8, 6020 Innsbruck, Austria
\and Astrophysics Group, Keele University, Staffordshire, ST5 5BG, United Kingdom 
\and Department of Astrophysics/IMAPP, Radboud University Nijmegen, P.O. Box 9010, 6500 GL Nijmegen, The Netherlands
\and South African Astronomical Observatory, P.O. Box 9, Observatory, 7935, South Africa
\and Southern African Large Telescope Foundation, P.O. Box 9, Observatory, 7935, South Africa
\and Las Cumbres Observatory Global Telescope Network, 6740 Cortona Drive, Suite 102, Santa Barbara, CA 93117, USA
\and Department of Physics, Broida Hall, University of California, Santa Barbara, CA 93106, USA
\and Division of Geological and Planetary Sciences, California Institute of Technology, Pasadena, CA 91125, USA
\and Nordic Optical Telescope, Apartado 474, 38700 Santa Cruz de La Palma, Spain
\and Institute of Astronomy, K.U.Leuven, Celestijnenlaan 200D, B-3001 Heverlee, Belgium
\and Australian Astronomical Observatory, PO Box 915, North Ryde NSW 1670, Australia
\and Centre of Astrophysics Research, University of Hertfordshire, College Lane, Hatfield AL10 9AB, United Kingdom}
\date{Received \ Accepted}

\abstract{Type Ia supernovae (SN Ia) are the most important standard candles for measuring the expansion history of the universe. The thermonuclear explosion of a white dwarf can explain their observed properties, but neither the progenitor systems nor any stellar remnants have been conclusively identified. Underluminous SN\,Ia have been proposed to originate from a so-called double-detonation of a white dwarf. After a critical amount of helium is deposited on the surface through accretion from a close companion, the helium is ignited causing a detonation wave that triggers the explosion of the white dwarf itself. We have discovered both shallow transits and eclipses in the tight binary system CD$-$30$^\circ$11223 composed of a carbon/oxygen white dwarf and a hot helium star, allowing us to determine its component masses and fundamental parameters. In the future the system will transfer mass from the helium star to the white dwarf. Modelling this process we find that the detonation in the accreted helium layer is sufficiently strong to trigger the explosion of the core. The helium star will then be ejected at so large a velocity that it will escape the Galaxy. The predicted properties of this remnant are an excellent match to the so-called hypervelocity star US\,708, a hot, helium-rich star moving at more than $750\,{\rm km\,s^{-1}}$, sufficient to leave the Galaxy. The identification of both progenitor and remnant provides a consistent picture of the formation and evolution of underluminous type Ia supernovae.
}

\titlerunning{Progenitor and remnant of SN\,Ia}

\maketitle

\section{Introduction}

The search for the progenitors of SN\,Ia is ongoing, but the observational evidence remains inconclusive. In the standard single-degenerate scenarios mass is transferred in a stable way by either a main sequence star or a red giant to a white dwarf (WD) companion. In the double-degenerate scenario a close binary consisting of two white dwarfs shrinks due to angular momentum lost by the emission of gravitational waves and eventually merges. Possible progenitor systems have been proposed for both channels, but not conclusively identified yet. Although most SN\,Ia form a homogeneous class, about one third of them differ significantly in their luminosities and other observational properties and their proper classification is crucial when using such events as standard candles for cosmology (Wang \& Han \cite{wang12}). 

Underluminous SN\,Ia have been proposed to originate from a so-called double-detonation of a white dwarf. After a critical amount of helium is deposited on the surface through accretion from a close companion, the helium is ignited causing a detonation wave that triggers the explosion of the white dwarf itself even if its mass is significantly lower than the Chandrasekhar limit (Nomoto \cite{nomoto82}; Woosley et al. \cite{woosley86}). Hydrodynamic simulations predict the explosion of a CO-WD with a minimum mass of only $\sim0.8\,M_{\rm \odot}$ as underluminous SN\,Ia triggered by the ignition of an He-shell of $\sim0.1\,M_{\rm \odot}$ (Fink et al. \cite{fink10}) in this so-called double-detonation scenario. He-stars have already been proposed as possible donors for the single-degenerate scenario (Yoon \& Langer \cite{yoon03}; Wang et al. \cite{wang09a,wang09b}) conveniently explaining the lack of hydrogen in the spectra of SN\,Ia. Recent studies indicate that this scenario might also be consistent with the lack of helium in standard SN\,Ia spectra as long as the accreted He-layer is thin (Sim et al. \cite{sim10}; Kromer et al. \cite{kromer10}).  

\begin{figure}[t!]
\begin{center}
 \includegraphics[width=9cm]{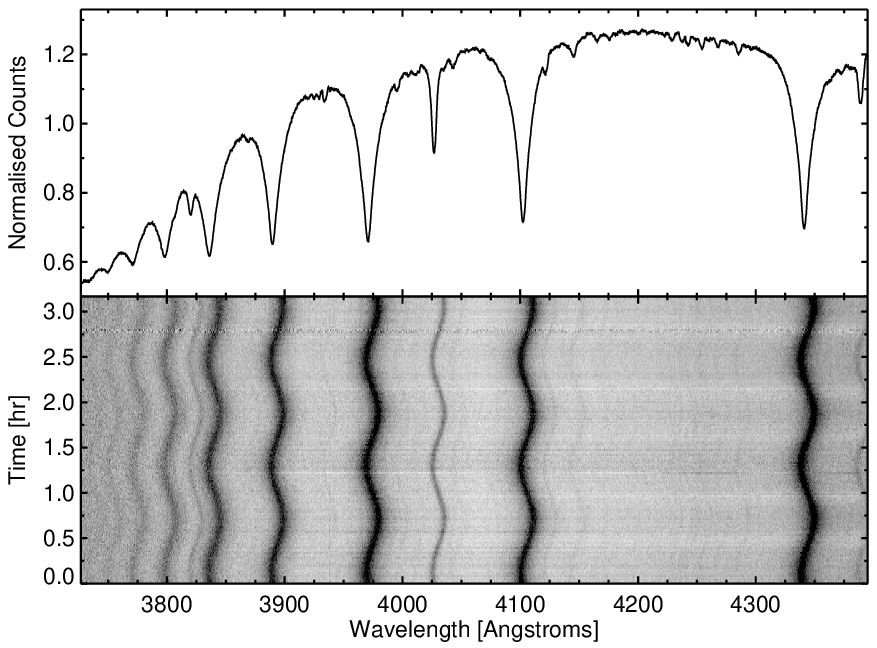}
 \includegraphics[width=9cm]{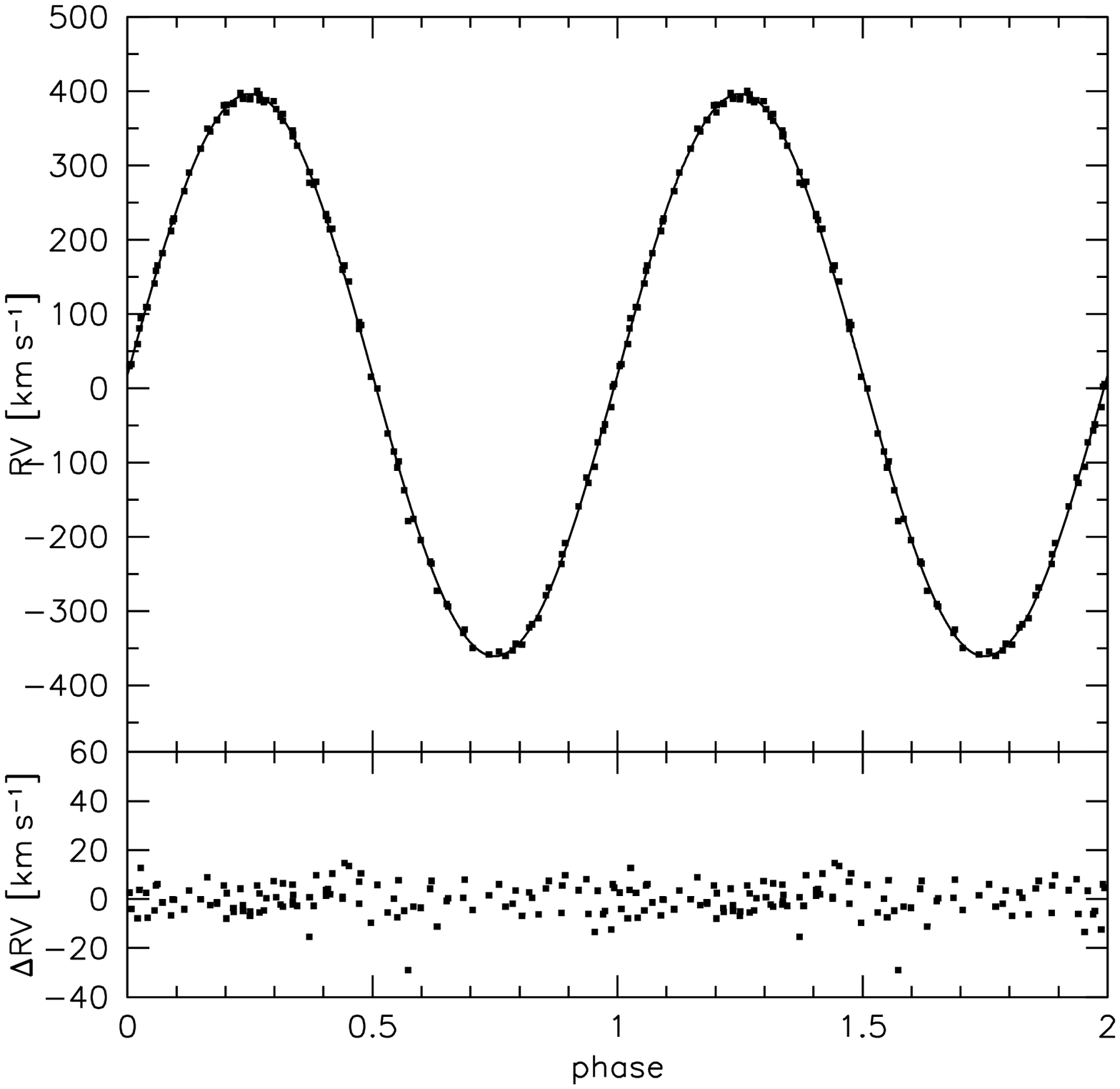}
 \caption{Upper panel: Spectrum of CD$-$30$^\circ$11223 coadded from 175 RV-corrected spectra taken with SOAR/Goodman. The hydrogen Balmer series is clearly visible as well as a prominent helium line at $4026\,{\rm \AA}$. The other features are rotationally broadened metal lines. Model spectra were matched to the hydrogen and helium lines to determine the atmospheric parameters of the hot subdwarf star. Middle panel: Trailed spectra taken with SOAR/Goodman and showing the short-period sinusoidal variations of the Doppler-shifted spectral lines caused by the motion of the visible sdB star. The close and compact white-dwarf companion of the subdwarf star is not visible in this dataset. Lower panel: Radial velocity curve of CD$-$30$^\circ$11223 derived from 105 spectra taken with WHT/ISIS plotted twice against orbital phase for better visualisation.}
 \label{spectrogram}
\end{center}
\end{figure}

In the course of the MUCHFUSS project (Geier et al. \cite{geier11a}), which aims at finding hot subdwarf binary systems with massive companions, we have discovered a possible progenitor for such a supernova consisting of a hot subdwarf B star (sdB) and a white dwarf in an extremely compact binary (Heber et al. \cite{heber13}; Geier et al. \cite{geier12}). This system, CD$-$30$^\circ$11223, was also independently discovered by Vennes et al. (\cite{vennes12}). 

Hot subdwarf stars are evolved, core helium-burning objects. About half of the sdB stars reside in close binaries with periods ranging from $\sim0.1\,{\rm d}$ to $\sim30\,{\rm d}$ (Maxted et al. \cite{maxted01}; Napiwotzki et al, \cite{napiwotzki04a}). The sdB is the core of a former red giant star that has been stripped off almost all of its hydrogen envelope through interaction with a companion star. The mass of the emerging sdB star is constrained to about half a solar mass in order to allow central helium burning. After the helium-burning phase the sdB star will turn into a white dwarf.   

Because the components' separation in these systems is much less than the size of the subdwarf progenitor in its red-giant phase, these systems must have experienced a common-envelope (CE) and spiral-in phase (Han et al. \cite{han02,han03} and references therein). In this scenario, a star evolving to become a red giant swallows a nearby companion. Due to friction the core of the giant and the more compact companion spiral towards each other in a common envelope. The orbital energy lost during this process is deposited in the envelope until it is eventually ejected leaving a close binary system as remnant. 

Although most of the close companions to sdB stars are low-mass main sequence stars, brown dwarfs or low-mass WDs ($\sim0.5\,M_{\rm \odot}$), more massive compact companions like WDs, neutron stars or black holes have been either observed or predicted by theory (Geier et al. \cite{geier07,geier10}). The short-period sdB+WD binary KPD\,1930$+$2752 is regarded as progenitor candidate for an SN\,Ia (Maxted et al. \cite{maxted00}; Geier et al. \cite{geier07}).

Here we report the discovery of both shallow transits and eclipses in the tight binary system CD$-$30$^\circ$11223 composed of a carbon/oxygen white dwarf and a hot helium star, allowing us to determine its component masses and fundamental parameters. This system turns out to be an excellent progenitor candidate for the double-detonation SN\,Ia scenario and can be linked to the hypervelocity subdwarf US\,708, the likely donor remnant of such an event.

\section{Observations}

CD$-$30$^\circ$11223 ($\alpha_{2000}=14^{\rm h}11^{\rm m}16{\stackrel{\rm s}{\displaystyle.}}2$, $\delta_{2000}=-30^{\rm \circ}53'03''$, $m_{\rm V}=12.3\,{\rm mag}$) was selected as UV-excess object and spectroscopically identified to be an sdB star (Vennes et al. \cite{vennes11}; N\'emeth et al. \cite{nemeth12}). We selected this star as a bright backup target for our MUCHFUSS follow-up campaign. Due to unfavourable observing conditions, which prevented us from observing our main targets, two medium resolution spectra ($R\sim2200,\lambda=4450-5110\,{\rm \AA}$) were taken consecutively with the EFOSC2 spectrograph mounted at the ESO\,NTT on June 10, 2012. The radial velocity shift between those two spectra turned out to be as high as $600\,{\rm km\,s^{-1}}$.

First spectroscopic follow-up data was obtained with the grating spectrograph mounted on the SAAO-1.9m telescope on July 2, 2012. The RV-curve derived from 18 single spectra confirmed the short orbital period of $0.0498\,{\rm d}$ and a high RV-semiamplitude ($K=370\pm14\,{\rm km\,s^{-1}}$). In order to improve the orbital solution and minimize the effect of orbital smearing, we took another 105 spectra with the ISIS spectrograph ($R\sim4000,\lambda=3440-5270\,{\rm \AA}, T_{\rm exp}=2\,{\rm min}$) mounted at the WHT during a dedicated MUCHFUSS follow-up run from July 9 to 12, 2012. Another 175 spectra were taken with the Goodman spectrograph mounted at the SOAR telescope ($R\sim7700, \lambda=3700-4400\,{\rm \AA}, T_{\rm exp}=1\,{\rm min}$) on July 16, 2012.

CD$-$30$^\circ$11223 was observed by the SuperWASP planetary transit survey (Pollacco et al. \cite{pollacco06}). The light curve contains $23\,678$ measurements taken from May 4, 2006 to August 1, 2011. 

$3.6\,{\rm hr}$ of time-series photometry in the V-band ($T_{\rm exp}=3\,{\rm s}$) were taken with SOAR/Goodman on July 6, 2012 under photometric conditions. The light curve was extracted using an aperture that minimizes the standard deviation of the two comparison stars used divided by each other at low airmass and a flat-field correction has been applied. The combination of short exposure times and bright stars make scintillation the dominant noise source for our photometry. Another source of noise is likely to be caused by a small scatter in effective integration time of the order of a few ms.

\section{Orbital and atmospheric parameters}

\begin{figure}[t!]
\begin{center}
 \includegraphics[width=8cm]{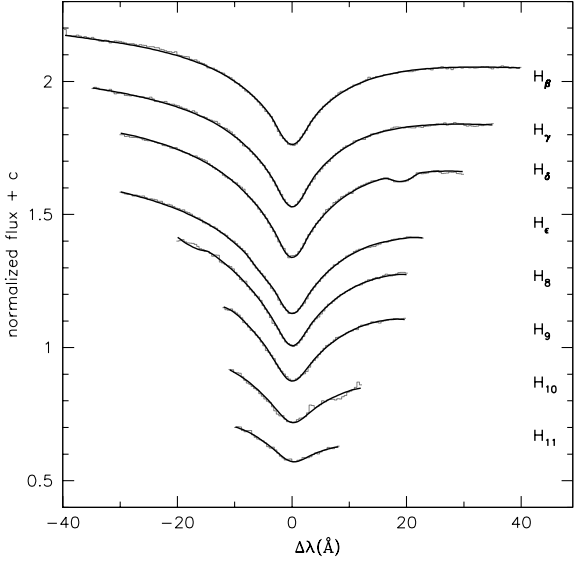}
 \caption{Fit of synthetic LTE models to the hydrogen Balmer lines of a coadded ISIS spectrum. The normalized fluxes of the single lines are shifted for better visualisation.}
 \label{fitH}
\end{center}
\end{figure}

The light curve shows variations caused by the ellipsoidal deformation of the sdB primary, which is triggered by the tidal influence of the compact companion, as well as Doppler boosting, caused by the extreme orbital motion of the sdB (Shakura \& Postnov \cite{shakura87}, see also Bloemen et al. \cite{bloemen11} and references therein). The ephemeris has been derived from the SWASP data based on fitting a harmonic series. Due to the timebase of more than five years the derived orbital period of $0.0489790724\pm0.0000000018\,{\rm d}$ is very accurate and perfectly consistent with the independent determination ($P=0.04897906\pm0.00000004\,{\rm d}$) by Vennes et al. (\cite{vennes12}). 

\begin{figure}[t!]
\begin{center}
 \includegraphics[width=8cm]{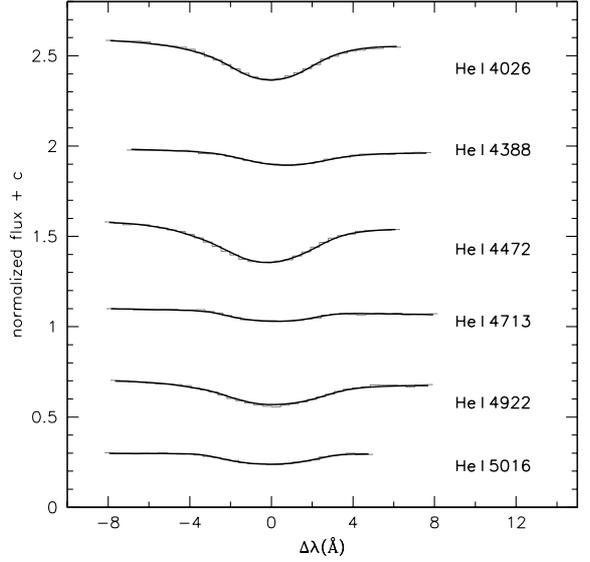}
 \caption{Fit of synthetic LTE models to the helium lines (see Fig~\ref{fitH}).}
 \label{fitHe}
\end{center}
\end{figure}

\begin{figure}
\begin{center}
 \includegraphics[width=8cm]{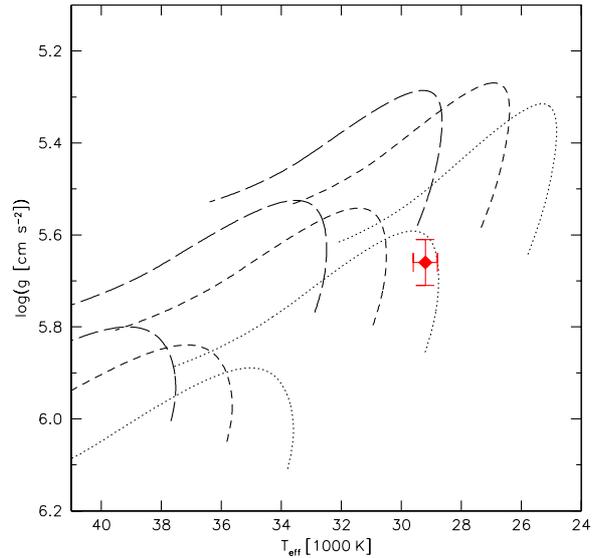}
 \caption{$T_{\rm eff}-\log{g}$ diagram. Evolutionary tracks (solar metallicity) of core helium-burning star with masses of $0.45\,M_{\rm \odot}$ (dotted lines), $0.50\,M_{\rm \odot}$ (short-dashed lines) and $0.55\,M_{\rm \odot}$ (long-dashed lines) are plotted for different hydrogen envelope masses ($0.000\,M_{\rm \odot}$, $0.001\,M_{\rm \odot}$, $0.005\,M_{\rm \odot}$ from the lower left to the upper right).$^{6}$ The diamond marks CD$-$30$^\circ$11223.}
 \label{tefflogg}
\end{center}
\end{figure}

\begin{figure*}[t!]
 \includegraphics[width=11cm]{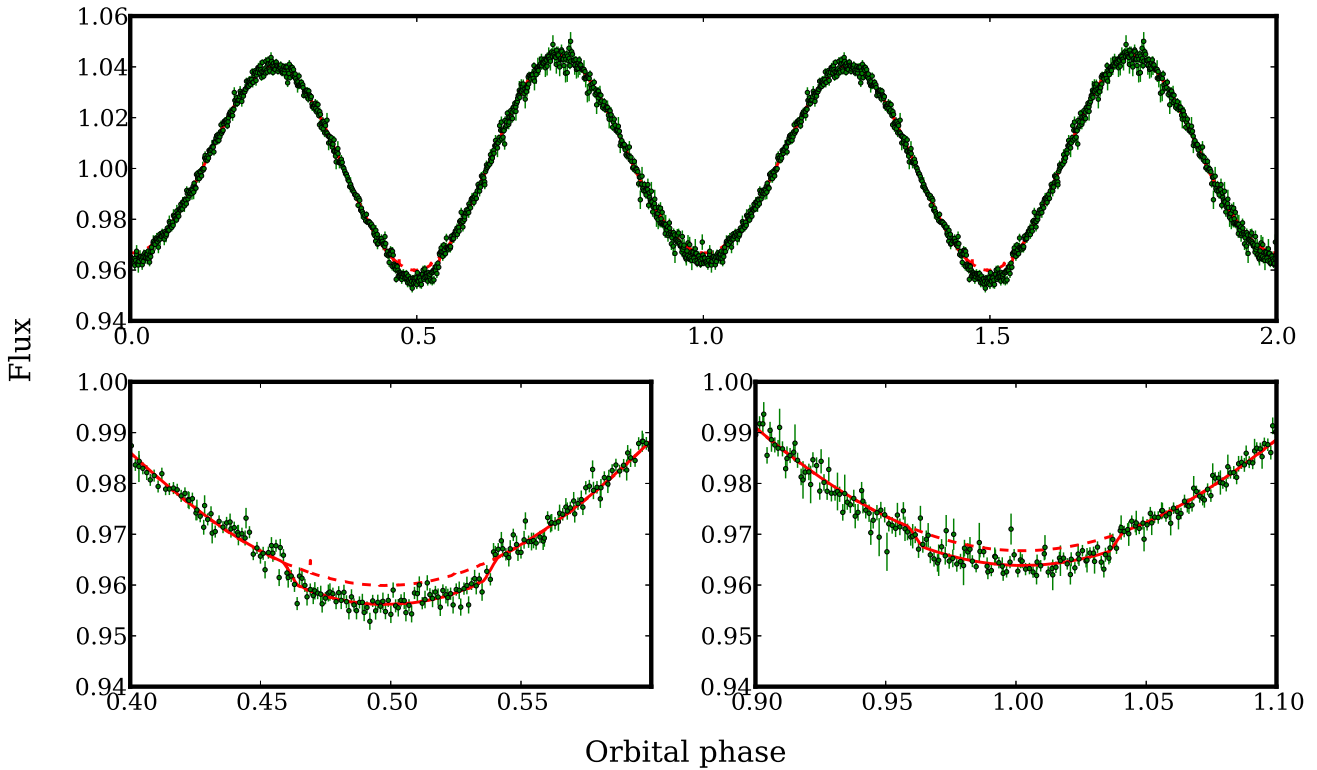}
 \caption{Upper panel: V-band light curve of CD$-$30$^\circ$11223 taken with SOAR/Goodman (green) with superimposed model (red) plotted twice against orbital phase for better visualisation. The dashed red curve marks the same model without transits and eclipses. The sinusoidal variation is caused by the ellipsoidal deformation of the hot subdwarf due to the tidal influence of the compact white dwarf. The difference in the maxima between phase $0.25$ and $0.75$ originates from the relativistic Doppler boosting effect, which is usually not detectable with ground-based telescopes. Lower panels: Close-up on the transit of the WD in front of the sdB (left). It is even possible to detect the eclipse of the WD by the sdB (right).}
 \label{lc_soar_fit}
\end{figure*}

The radial velocities were measured by fitting a set of mathematical functions to the hydrogen Balmer lines as well as helium lines using the FITSB2 routine (Napiwotzki et al. \cite{napiwotzki04b}). Three functions (Gaussians, Lorentzians and polynomials) are used to match the continuum, the line and the line core, respectively and mimic the typical Voigt profile of spectral lines. The profiles are fitted to all suitable lines simultaneously using $\chi^{2}$-minimization and the RV shift with respect to the rest wavelengths is measured. Assuming circular orbits sine curves were fitted to the RV data points in fine steps over a range of test periods. The two datasets obtained with ISIS and Goodman are treated separately to investigate systematic errors. Details about the analysis method and error estimation are given in (Geier et al. \cite{geier11b}). The derived orbital parameters from the ISIS dataset ($K=378.6\pm1.0\,{\rm km\,s^{-1}}$, $\gamma=17.6\pm0.7\,{\rm km\,s^{-1}}$, see Fig.~\ref{spectrogram}, lower panel) and the Goodman dataset ($K=374.5\pm1.1\,{\rm km\,s^{-1}}$, $\gamma=21.3\pm0.8\,{\rm km\,s^{-1}}$, see Fig.~\ref{spectrogram}, upper panel) are consistent taking into account that systematic uncertainties are usually somewhat higher than the statistical $1\,\sigma$ errors given here. The deviation in system velocity is most likely caused by a slight systematic zero-point shift between the two instruments. For further analysis we used the average values. Those values are in reasonable agreement with the results ($K=386.9\pm1.9\,{\rm km\,s^{-1}}$, $\gamma=31.5\pm1.3\,{\rm km\,s^{-1}}$) of Vennes et al. (\cite{vennes12}), although somewhat discrepant.

The atmospheric parameters effective temperature $T_{\rm eff}$, surface gravity $\log\,g$, helium abundance and projected rotational velocity were determined by fitting simultaneously the observed hydrogen and helium lines of the single spectra with metal-line-blanketed LTE model spectra (Heber et al. \cite{heber00}) as described in Geier et al. (\cite{geier07}). No significant variations of the parameters with orbital phase have been detected. Average values and standard deviations have been calculated for the ISIS ($T_{\rm eff}=28800\pm200\,{\rm K}$, $\log{g}=5.67\pm0.03$, $\log{y}=-1.50\pm0.07$, $v_{\rm rot}\sin{i}=180\pm8\,{\rm km\,s^{-1}}$) and Goodman datasets ($T_{\rm eff}=29600\pm300\,{\rm K}$, $\log{g}=5.65\pm0.05$, $\log{y}=-1.46\pm0.14$, $v_{\rm rot}\sin{i}=174\pm12\,{\rm km\,s^{-1}}$), separately. We adopt the average values from both datasets for further analysis. The final helium abundance is taken from the ISIS data, because of the higher number of He-lines in the spectral range. 

The derived parameters are consistent with literature values within the uncertainties (Vennes et al. \cite{vennes11}; N\'emeth et al. \cite{nemeth12}). More detailed information about the systematic errors of this method can be found in Geier et al. (\cite{geier07,geier11b}). Table~1 shows the orbital and atmospheric parameters, Fig.~\ref{tefflogg} the position of CD$-$30$^\circ$11223 in the $T_{\rm eff}-\log{g}$ diagram.

\section{Light curve analysis}

The light curve obtained with SOAR/Goodman was analysed by fitting models calculated with the {\sc lcurve} code written by TRM (see Fig.~\ref{lc_soar_fit}, Copperwheat et al. \cite{copperwheat10}). The code uses grids of points modelling the two stars and takes into account limb darkening, gravity darkening, mutual illumination effects, Doppler boosting and gravitational lensing. Since the masses and radii of both components are strongly correlated, those parameters have been constrained using Markov chain Monte Carlo simulations. A detailed description of the analysis method is given by Bloemen et al. (\cite{bloemen11}).

\begin{figure*}
 \includegraphics[width=10cm]{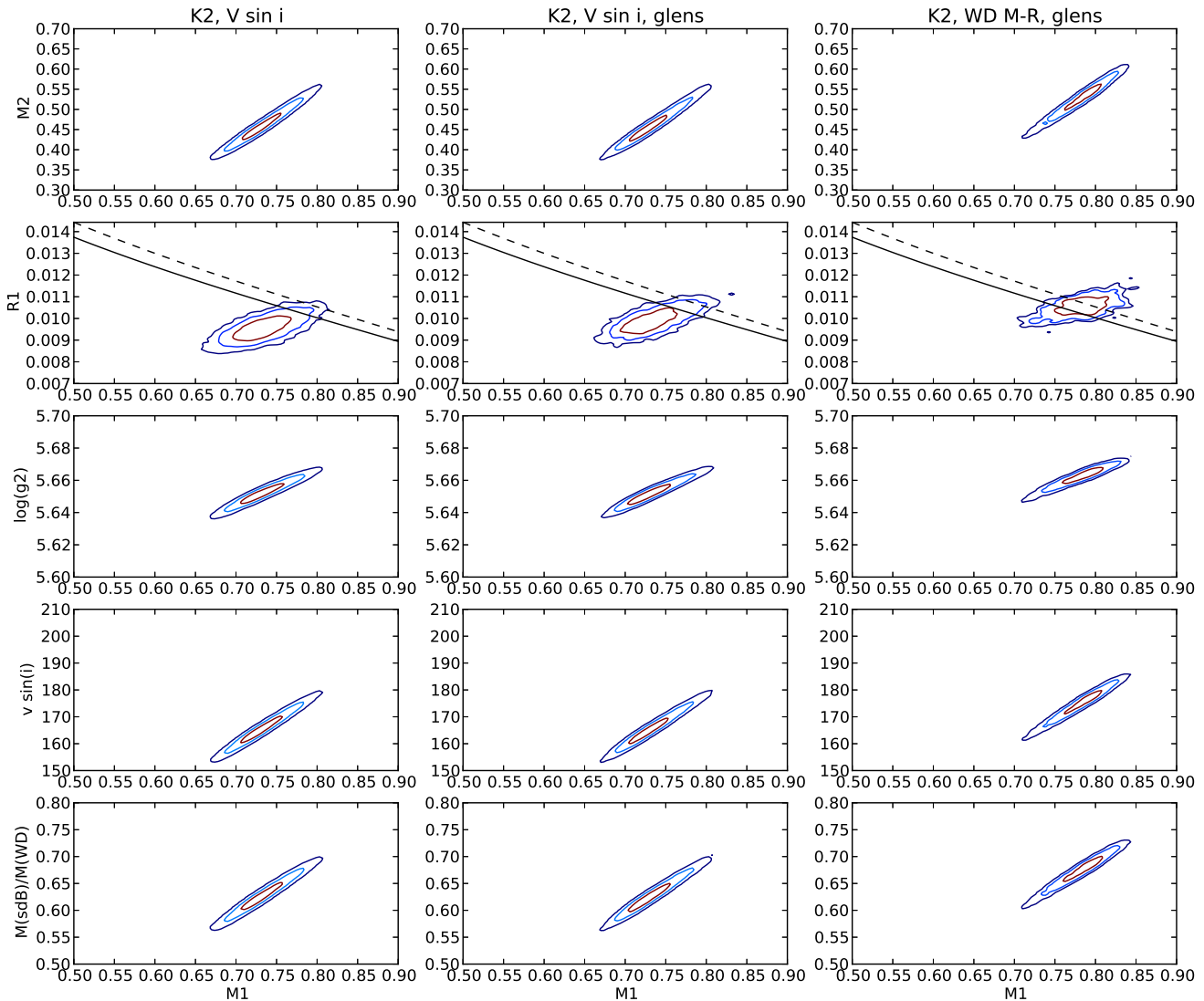}
 \caption{Parameters of the CD$-$30$^\circ$11223 system plotted against the mass of the WD companion with contours marking the $1\,\sigma$ (red), $2\,\sigma$ (light blue), and $3\,\sigma$ (dark blue) levels of confidence. In the left column tidal synchronisation of the sdB primary is assumend, while in the middle column the effect of microlensing has been taken into account in addition. In the right column the WD mass-radius relation has been used as additional constraint instead. The solid lines in the mass-radius plots (second panels from above) mark the zero-temperature mass-radius relation for WDs, the dashed lines more realistic models with $5\%$ inflation. Note that both the $v_{\rm rot}\sin{i}$ and the $\log{g}$ derived from the light curve are in agreement with the values derived from spectroscopy.}
 \label{M1M2R1R2}
\end{figure*}

In order to determine masses and radii of both the sdB and the WD, we used two different prior constraints. In each case the $K$ derived from spectroscopy is used. First, we assumed tidal synchronisation of the sdB primary (solution 1), which is a reasonable assumption given the short orbital period of the system. The $v_{\rm rot}\sin{i}$ derived from spectroscopy is measured making the simplified assumption of a spherical, linear limb darkened star (limb darkening coefficient $0.3$). In order to take into account the additional effects of limb darkening and gravitational darkening of the Roche-distorted star we calculated a correction factor of $0.963$ by comparing the slightly different line profiles calculated under both assumptions (Claret \& Bloemen \cite{claret11}). This correction was applied to the $v_{\rm rot}\sin{i}$ before deriving the binary parameters. The best fit is achieved for an sdB mass of $0.47\pm0.03\,M_{\rm \odot}$ and a WD mass of $0.74\pm0.02\,M_{\rm \odot}$. However, we found that the radius of the WD is about $10\%$ smaller than predicted by the zero-temperature mass-radius relation for WDs, which provides a lower limit for the WD radius (see Fig.~\ref{M1M2R1R2} second panel from above, Verbunt \& Rappaport \cite{verbunt88}).

In order to explore the influence of this discrepancy, we imposed the restriction that the white dwarf is within $2\%$ of the $M$-$R$-relation and allowed for deviations from corotation (solution 2). We determined an estimate for the temperature of the WD from our light curve analysis. Since we see a significant feature when the sdB occults the WD, we can derive a black-body temperature of $24700\pm1200\,{\rm K}$ for the WD, which leads to a radius about $5\%$ higher than expected from the zero-temperature relation. We adopt this more realistic value for our analysis. In this case the derived masses are somewhat higher ($M_{\rm sdB}=0.54\pm0.02\,M_{\rm \odot}$, $M_{\rm WD}=0.79\pm0.01\,M_{\rm \odot}$). 

Although both solutions are consistent within their uncertainties we refrain from favouring one over the other. In order to calculate the kinematics and the further evolution of the system, we adopt the average values for the component masses ($M_{\rm sdB}=0.51\,M_{\rm \odot}$, $M_{\rm WD}=0.76\,M_{\rm \odot}$, see Fig.~\ref{M1M2R1R2}).

Comparing the derived sdB masses with evolutionary tracks for core helium-burning objects (Fig.~\ref{tefflogg}), it can be seen that the appropriate tracks are consistent with the position of CD$-$30$^\circ$11223 in the $T_{\rm eff}-\log{g}$ diagram. Furthermore, the effective temperature and surface gravity of the star tell us that the sdB has just recently been formed and started the core helium-burning phase, which typically lasts for about $100\,{\rm Myr}$.

\begin{table*}
\label{compmasses}
\caption{Parameters of the CD$-$30$^\circ$11223 system}
\begin{center}
\begin{tabular}{llll}

\hline
\noalign{\smallskip}

Visual magnitude$^{\dag}$ & $m_{\rm V}$ & [mag] & $12.342\pm0.003$ \\
Proper motion$^{\ddag}$                 & $\mu_{\rm \alpha}\cos{\delta}$ & [mas/yr] & $9.5\pm2.2$ \\
				     & $\mu_{\rm \delta}$ & [mas/yr] & $-5.6\pm2.2$ \\
\noalign{\smallskip}
\hline
\noalign{\smallskip}

Atmospheric parameters of the subdwarf& & & \\

\noalign{\smallskip}
\hline
\noalign{\smallskip}

Effective temperature & $T_{\rm eff}$ & [K] & $29\,200\pm400$\\
Surface gravity & $\log{g}$           & & $5.66\pm0.05$\\
Helium abundance & $\log{y}$          & & $-1.50\pm0.07$\\
Projected rotational velocity & $v_{\rm rot}\sin{i}$ & [${\rm km\,s^{-1}}$] & $177\pm10$\\

\noalign{\smallskip}
\hline
\noalign{\smallskip}

Orbital parameters & & & \\

\noalign{\smallskip}
\hline
\noalign{\smallskip}
               & $T_{\rm 0}$ & [BJD UTC] & $2455113.205908\pm0.000363$ \\
Orbital period & $P$ & [d] & $0.0489790724\pm0.0000000018$\\
RV semi-amplitude & $K$ & [${\rm km\,s^{-1}}$] & $376.6\pm1.0$\\
System velocity & $\gamma$ & [${\rm km\,s^{-1}}$] & $19.5\pm2.0$\\
Binary mass function & $f(M)$ & [$M_{\rm \odot}$] & $0.271\pm0.002$\\

\noalign{\smallskip}
\hline
\noalign{\smallskip}

Derived parameters & & & \\

\noalign{\smallskip}
\hline
\noalign{\smallskip}

Solution 1 & & & \\

\noalign{\smallskip}
\hline
\noalign{\smallskip}

sdB mass & $M_{\rm sdB}$ & [$M_{\rm \odot}$] & $0.47\pm0.03$\\
sdB radius & $R_{\rm sdB}$ & [$R_{\rm \odot}$] & $0.169\pm0.005$\\
WD mass & $M_{\rm WD}$ & [$M_{\rm \odot}$] & $0.74\pm0.02$\\
WD radius & $R_{\rm WD}$ & [$R_{\rm \odot}$] & $0.0100\pm0.0004$\\
Orbital inclination & $i$ & [$^{\rm \circ}$] & $83.8\pm0.6$ \\                        
Separation & $a$ & [$R_{\rm \odot}$] & $0.599\pm0.009$\\
Mass ratio & $q$ & & $0.63\pm0.02$\\

\noalign{\smallskip}
\hline
\noalign{\smallskip}

Solution 2 & & & \\

\noalign{\smallskip}
\hline
\noalign{\smallskip}

sdB mass & $M_{\rm sdB}$ & [$M_{\rm \odot}$] & $0.54\pm0.02$\\
sdB radius & $R_{\rm sdB}$ & [$R_{\rm \odot}$] & $0.179\pm0.003$\\
WD mass & $M_{\rm WD}$ & [$M_{\rm \odot}$] & $0.79\pm0.01$\\
WD radius & $R_{\rm WD}$ & [$R_{\rm \odot}$] & $0.0106\pm0.0002$\\
Orbital inclination & $i$ & [$^{\rm \circ}$] & $82.9\pm0.4$ \\                        
Separation & $a$ & [$R_{\rm \odot}$] & $0.619\pm0.005$\\
Mass ratio & $q$ & & $0.68\pm0.01$\\

\hline\\
\end{tabular}
\end{center}
$^{\dag}$The visual magnitude is taken from Vennes et al. (\cite{vennes12})\\
$^{\ddag}$Proper motions taken from Roeser et al. (\cite{roeser10}).\\
\end{table*}

\section{Gravitational wave radiation}

Because of its short orbital period, CD$-$30$^\circ$11223 is expected to be a strong source of gravitational waves. We therefore calculated the current gravitational wave emission of CD$-$30$^\circ$11223. The gravitational wave strain amplitude $h$ scales with the masses of both binary components, the binary inclination, the orbital period and the distance of the system. 

We calculate it as described in Roelofs et al. (\cite{roelofs07}) to be as high as $\log{h}=-21.5\pm0.3$. CD$-$30$^\circ$11223 should therefore be one of the strongest gravitational wave sources detectable with missions like NGO/eLISA (Kilic et al. \cite{kilic12}; Nelemans \cite{nelemans09}). It sticks out even further, because due to the presence of eclipses its binary parameters are determined to very high accuracy. Therefore this system can be used as verification source for upcoming space missions.

No period change due to the orbital shrinkage caused by the emission of gravitational wave radiation has been detected in the SWASP data ($\dot P=1.01\times10^{-12}\pm3.38\times10^{-12}\,{\rm s\,s^{-1}}$). This non-detection is consistent with the theoretically expected value of $\dot P\sim6\times10^{-13}\,{\rm s\,s^{-1}}$. However, within only a few more years the orbital shrinkage should become detectable.

\section{Binary evolution calculations}

\begin{figure*}[t!]
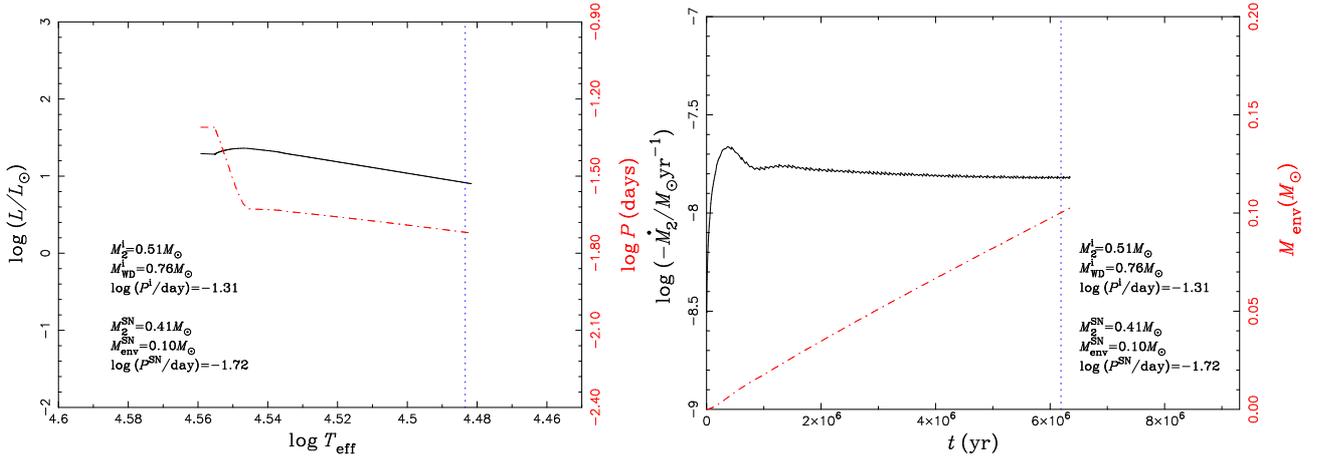

   \begin{center}
\includegraphics[width=6cm,angle=270]{f2.ps}
\includegraphics[width=6cm,angle=270]{f1.ps}
 \caption{Results of binary evolution calculations with initial masses of the two components and the orbital period similar to the sdB+WD binary system. Left panel: The evolutionary track of the He star is shown as solid curve and the evolution of the orbital period as dash-dotted curve. Right panel: The solid and dash-dotted curves show the mass-transfer rate and the mass of the WD envelope (He shell) varying with time after the He star fills its Roche lobe, respectively. Dotted vertical lines in both panels indicate the position where the double-detonation may happen (the mass of the He shell increases to $\sim0.1\,M_{\odot}$). The initial binary parameters and the parameters at the moment of the SN explosion are also given in the two panels.}
\label{masstransfer}
   \end{center}
\end{figure*}

More interesting than the present state of this system is its future evolution, which can now be studied in detail using theoretical models.
Employing Eggleton's stellar evolution code (Eggleton \cite{eggleton71,eggleton72,eggleton73}),
we calculate the evolution of the sdB star and its WD
companion. The code has been updated
with the latest input physics over the past four decades (Han et al. \cite{han94}; Pols et al. \cite{pols95,pols98}).
Roche lobe overflow (RLOF) is
treated within the code described by Han et al. (\cite{han00}).
We set the ratio of mixing length to
local pressure scale height, $\alpha=l/H_{\rm p}$, to be 2.0. We assume that the binary model
starts with a 0.51\,$M_{\rm \odot}$ He star and a 0.76$\,M_{\rm \odot}$ CO WD
having a 0.049\,d orbit period, similar to the initial model of the sdB star and its WD companion.
Additionally, orbital angular momentum loss due to gravitational wave
radiation is included by adopting a standard formula presented
by Landau \& Lifshitz (\cite{landau71}),
\begin{equation}
{d\,\ln J_{\rm GR}\over dt} = -{32G^3\over 5c^5}\,{M_{\rm WD} M_2
(M_{\rm WD}+M_2)\over a^4},
\end{equation}
where $G$, $c$, $M_{\rm WD}$ and $M_2$ are the gravitational
constant, vacuum speed of light, the mass of the accreting WD and
the mass of the companion sdB star, respectively.

In the He double-detonation model, if the mass-accretion rate 
is higher than $4\times10^{-8}\,M_{\rm \odot}\,\mathrm{yr}^{-1}$, 
the WD can increase its mass (Woosley et al. \cite{woosley86}; Wang et al. \cite{wang09a}). However, for low mass-accretion
rates ($1\times10^{-9}\,M_{\rm \odot}\mathrm{yr}^{-1}\lesssim|\dot M_2|\lesssim4\times10^{-8}\,M_{\rm \odot}\mathrm{yr}^{-1}$),
compressional heating at the base of the accreted He layer plays
no significant role, and a layer of unburned He can be accumulated
on the surface of the WD. For the case with a slower mass-accretion rate ($|\dot M_2|<1\times10^{-9}\,M_{\odot} \mathrm{yr}^{-1}$), the He flash is strong enough to form a He
detonation but too weak to initiate a carbon detonation, resulting in that only a single He detonation
wave propagates outward (Nomoto \cite{nomoto82}). 

We assume that if such a CO WD accumulating He enters this `low' accretion rate regime and
accumulates $0.1\,M_{\rm \odot}$ of He on its surface, a detonation is
initiated at the base of the He shell layer (Fink et al. \cite{fink10}). Consequently, a
detonation in the core of the CO WD is presumed to follow, in which a sub-$M_{\rm Ch}$ SN Ia takes place.
In this model, only accreting WDs with a total mass (CO core $0.8~M_{\rm \odot}$ + helium shell $0.1~M_{\rm \odot}$) ${\geq} 0.9~M_{\odot}$ are
considered to result in potential sub-$M_{\rm Ch}$ SNe Ia,
since the CO-core with lower mass may not detonate and it is
unlikely to produce enough radioactive nickel observed in SNe Ia (Kromer et al. \cite{kromer10}).

The binary system starts with ($M_2^{\rm i}$, $M_{\rm WD}^{\rm
i}$, $\log (P^{\rm i}/{\rm d})$) $=$ (0.510, 0.760, $-$1.310), where
$M_2^{\rm i}$, $M_{\rm WD}^{\rm i}$ are the initial masses of the He star
star and of the CO WD in solar mass, and $P^{\rm i}$ is the
initial orbital period in days. Fig.~\ref{masstransfer} (left panel) shows the evolutionary
track of the He star and the evolution of the orbital period. Fig.~\ref{masstransfer} (right panel) 
displays the mass-transfer rate and the mass of the WD envelope varying with
time after the He star fills its Roche lobe.

Due to the short initial orbital period (0.049\,d) of the system,
angular momentum loss induced by gravitational wave radiation is large. This
leads to the rapid shrinking of the orbital separation.
After about 36\,million years, the He star begins to fill its
Roche lobe while it is still in the core helium-burning stage.
The mass-transfer rate is stable and at a low rate between
$1.6\times10^{-8}\,M_{\rm \odot}\mathrm{yr}^{-1}$ and $2.2\times10^{-8}\,M_{\rm \odot}\mathrm{yr}^{-1}$, resulting in the formation of
a He shell on the surface of the CO WD. After about 6\,million
years, the mass of the He shell increases to $\sim0.1\,M_{\rm \odot}$
in which a double-detonation may happen at the base of the He shell layer.
At this moment, the mass of the He star is
$M^{\rm SN}_2=0.41\,M_{\rm \odot}$ and the orbital period is $\log(P^{\rm SN}/{\rm d})=-1.72$ ($P=0.019\,{\rm d}$).

\section{Hypervelocity sdO as donor remnant}

Theoretical predictions about whether or not a progenitor candidate will explode as SN\,Ia are useful, but in general difficult to test. Usually the theoretically predicted SN rates are compared to the observed ones, but these comparisons are often hampered by selection effects. A more direct proof would be the identification of the remnant objects. We therefore follow the future evolution of CD$-$30$^\circ$11223. At the end of the He-accretion phase and just before the SN event, the orbital period of the binary is predicted to have shrunk to $0.019\,{\rm d}$ due to the further loss of orbital energy through the emission of gravitational waves. The sdB primary lost a fair amount of mass ($\sim0.1\,M_{\rm \odot}$), which was transferred to the WD companion. The orbital velocity of the sdB will be about $600\,{\rm km\,s^{-1}}$ and therefore close to the Galactic escape velocity. As soon as the WD is disrupted, the sdB will be ejected. Depending on the ejection direction of such an object relative to its trajectory around the Galactic centre, the Galactic rest frame velocity could be even higher by up to $240\,{\rm km\,s^{-1}}$. In this case the remnant star will leave the Galaxy.

Such so-called hypervelocity stars have indeed been discovered (Brown et al. \cite{brown05}; Hirsch et al. \cite{hirsch05}; Edelmann et al. \cite{edelmann05}). However, all but one of the known 22 objects are intermediate-mass main-sequence star. This enigmatic star (US\,708) has been classified as helium-rich hot subdwarf travelling at a Galactic rest frame velocity of at least $750\,{\rm km\,s^{-1}}$ (Hirsch et al. \cite{hirsch05}), which matches the predicted ejection velocity of CD$-$30$^\circ$11223 very well. It was proposed that this star might be the ejected He-donor after the WD companion exploded as SN\,Ia (Justham et al. \cite{justham09}; Wang \& Han \cite{wang09c}). 

In this scenario, the compact binary CD$-$30$^\circ$11223 and the hypervelocity star US\,708 represent two different stages of an evolutionary sequence linked by a SN\,Ia explosion. The existence of objects like US\,708 thus provides evidence that binaries like CD$-$30$^\circ$11223 are viable SN\,Ia progenitor candidates. 

\section{Age of the binary system}

The analysis of our data also allows us to constrain the initial component masses and the age of the binary. Furthermore we can constrain both its past and future trajectory. 

\subsection{Kinematic analysis}

\begin{figure}[t!]
   \begin{center}
\includegraphics[width=9cm]{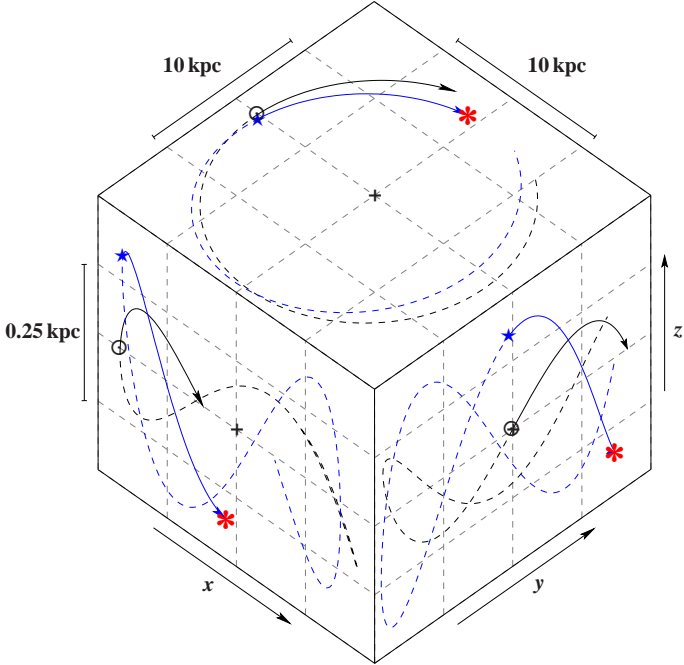}
 \caption{Three-dimensional trajectory of CD$-$30$^\circ$11223 in a Cartesian Galactic coordinate system with the z-axis pointing to the North Galactic pole. Current positions of CD$-$30$^\circ$11223 (blue $\star$), Sun (black $\odot$), and Galactic centre (black $+$) are marked. The approximate point in time of the supernova explosion is symbolized by the red asterisk, while the arrow marks the position of the Sun at that time. Solid lines indicate the future 42\,Myr, dashed lines the past 150\,Myr. CD$-$30$^\circ$11223's kinematic properties are obviously those of the local thin disc population.}
\label{kinematics}
   \end{center}
\end{figure}

Using a standard Galactic gravitational potential with a Sun-Galactic centre distance of 8.4\,kpc and a local standard of rest circular motion of $242\,{\rm km\,s^{-1}}$ (see Model I in Irrgang et al. \cite{irrgang13}), we computed the past and future trajectory of CD$-$30$^\circ$11223 (see Fig.~\ref{kinematics}). The orbit shows the typical characteristics of the local thin disc population, i.e., almost circular motion around the Galactic centre and small oscillations in direction perpendicular to the Galactic disc. The heliocentric distance to the star increases during the next 42\,Myr until the supernova is predicted to explode from its current value of $364\pm31\,{\rm pc}$ to about $1920\pm160\,{\rm pc}$.

CD$-$30$^\circ$11223 is by far the closest known SN\,Ia progenitor with respect to Earth. The explosion will take place in a direction of the sky close to the current positions of the constellations Ara and Norma. Adopting an absolute visual magnitude of up to $-19\,{\rm mag}$ for the SN\,Ia, the apparent magnitude seen from Earth might be as high as $\sim-7.6\,{\rm mag}$ or about as bright as SN\,1006, the brightest stellar event in recorded history so far (Winkler et al. \cite{winkler03}).

\begin{figure}[t!]
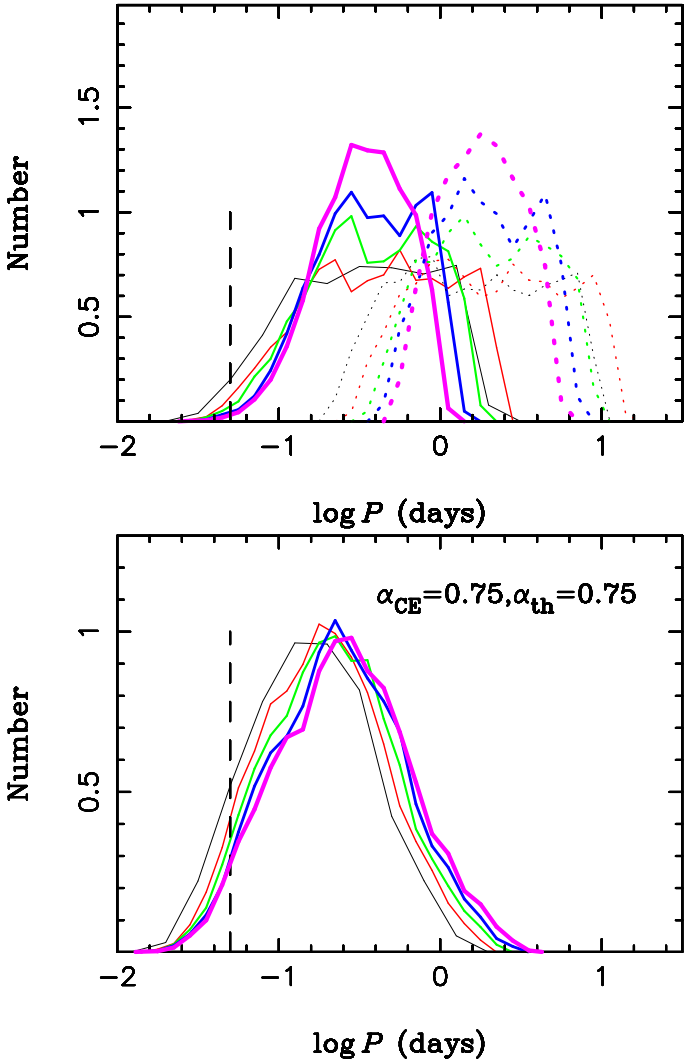

   \begin{center}
\includegraphics[width=7cm,angle=270]{xuefei-fig1.ps}
\includegraphics[width=7cm,angle=270]{xuefei-fig2.ps}
 \caption{Top panel: Orbital period distribution of sdB+WD binaries from the 2nd CE channel. Different colours mark different WD masses (black $0.6\,M_{\rm \odot}$, red $0.8\,M_{\rm \odot}$, green $1.0\,M_{\rm \odot}$, blue $1.2\,M_{\rm \odot}$, purple $1.4\,M_{\rm \odot}$). The vertical dashed line marks $P=0.05\,{\rm d}$. The contribution of thermal energy has not been included ($\alpha_{\rm th}=0$). The solid lines are for $\alpha_{\rm CE}=0.3$ and the dotted ones for $\alpha_{\rm CE}=1.0$. Bottom panel: Orbital period distribution of sdB+WD binaries from the CE channel involving more massive main sequence stars (i.e. $>2\,M_{\rm \odot}$). The values of $\alpha_{\rm CE}$ and  $\alpha_{\rm th}$ are indicated in the figure (see Fig.~\ref{Pdistrib_tip}).}
\label{Pdistrib_tip}
   \end{center}
\end{figure}

\subsection{Binary formation scenario}

CD$-$30$^\circ$11223 is the closest sdB binary known so far and the mass of its WD companion is higher than the average mass of CO WDs ($\sim0.6\,M_{\rm \odot}$). In order to explore the formation of this exceptional system, we performed a binary population synthesis study in a similar way as described by Han et al. (\cite{han02,han03}). For given WD masses ranging from $0.6\,M_{\rm \odot}$ to $1.4\,M_{\rm \odot}$, an initial set of $10^{6}$ WD+MS binaries was generated. For the main sequence stars the initial mass function of Salpeter was used. The orbital period distribution was assumed to be flat in $\log{a}$. The binaries have been evolved through the common envelope phase for different values of the CE-efficiency parameters $\alpha_{\rm CE}$, which is the fraction of the available orbital energy used to eject the envelope, and $\alpha_{\rm th}$, the contributed fraction of internal energy. 

In the standard scenario, which is called the 2nd CE channel, the progenitor of the sdB is a main sequence star of about solar mass and the common envelope is ejected right at the tip of the first giant branch (FGB, Han et al. \cite{han02}). However, this channel is not feasible to form binaries as close as CD$-$30$^\circ$11223, as shown in Fig.~\ref{Pdistrib_tip} (upper panel). The envelope at the tip of the FGB has a very low binding energy and can be ejected easily in the following CE. Thus, the orbital shrinkage during CE evolution is not significant and the produced sdB+WD system generally has an orbital period much longer than that of CD$-$30$^\circ$11223. Only for a very small value of $\alpha_{\rm CE}=0.3$, which is very unlikely, some binaries reach the margin of $0.05\,{\rm d}$. Indeed, the median period of the observed sdB binaries is as high as $\sim0.6\,{\rm d}$ (Geier et al. \cite{geier11a}).

However, an sdB+WD binary can also be formed when the main-sequence progenitor of the subdwarf has an initial mass larger than $2\,M_{\rm \odot}$ and fills its Roche lobe during the Hertzsprung Gap or at the base of the FGB. In this case, the envelope is more tightly bound and the orbital shrinkage required to eject the CE becomes higher. In Fig.~\ref{Pdistrib_tip} (lower panel) the orbital period distribution is shown for this scenario when $\alpha_{\rm CE}=\alpha_{\rm th}=0.75$, similar to the best fitting model of Han et al. (\cite{han03}). As seen in the figure, short orbital periods just as in the case of CD$-$30$^\circ$11223 are expected. 

\begin{figure}[t!]
   \begin{center}
\includegraphics[width=7cm,angle=270]{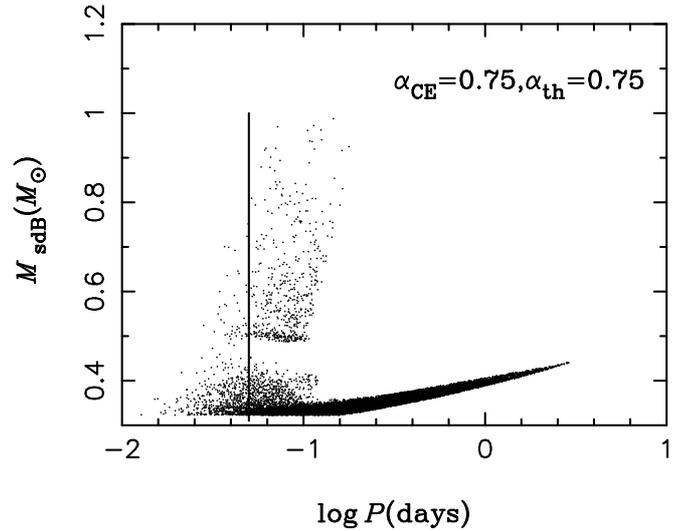}
 \caption{sdB mass is plotted over the orbital period assuming the mass of the WD to be $1.2\,M_{\rm \odot}$. The main-sequence progenitors of the sdBs have initial masses larger than $2\,M_{\rm \odot}$. The values of $\alpha_{\rm CE}$ and $\alpha_{\rm th}$ are indicated in the figure.}
\label{Pmass_th0}
   \end{center}
\end{figure}

Additional to the orbital period distribution, we also investigated the distribution of sdB masses formed via this channel. While the standard CE-scenario predicts a mass distribution with a sharp peak at $0.47\,M_{\rm \odot}$, the sdB masses from more massive main-sequence stars (i.e. $>2\,M_{\rm \odot}$) show a significant scatter for higher values of $\alpha_{\rm CE}$ and even more so, if we allow for a contribution of thermal energy in the CE-process by increasing the parameter $\alpha_{\rm th}$. The sdB mass for this channel largely depends on the mass of the progenitor and can range from $0.3$ to $1.0\,M_{\rm \odot}$ (see Fig.~\ref{Pmass_th0}). This is consistent with the sdB mass of up to $0.54\,M_{\rm \odot}$ determined in the case of CD$-$30$^\circ$11223. 

We therefore conclude that CD$-$30$^\circ$11223 was most likely formed via CE-ejection of a main sequence star with a mass larger than $2\,M_{\rm \odot}$, which means that it originated from a young stellar population.

\subsection{White dwarf cooling age and progenitor masses}

We derived a mass of $0.76\,M_{\rm \odot}$ for the WD companion based on observations. Using an initial-to-final mass relation for isolated WDs the mass of the progenitor should have ranged from $3$ to $4\,M_{\rm \odot}$ (see Fig.~7 in Kovetz et al. (\cite{kovetz09}) and references therein). Binary evolution is expected to lower the mass of the final WD and this progenitor mass estimate therefore has to be considered as lower limit. 

Assuming a lifetime on the main sequence $\tau_{\rm MS}=10^{10}\,{\rm yr}\times[M/M_{\rm \odot}]^{-2.5}$ the progenitor lived for a maximum of $640\,{\rm Myr}$. We constrain the temperature of the WD {in our light curve analysis} to be $\sim25000\,{\rm K}$. Therefore its cooling age is $\sim30-40\,{\rm Myr}$\footnote{Taken from cooling models for mixed C/O composition and for standard thick H and He layers, http://www.astro.umontreal.ca/~bergeron/CoolingModels/, Fontaine et al. \cite{fontaine01}}. The lifetime of the sdB on the extreme horizontal branch is of the same order as the one of the WD and we therefore derive a similar progenitor mass of more than $3\,M_{\rm \odot}$ consistent with the disc kinematics and the constraints from binary formation scenarios. 

\section{Conclusions}

Systems like CD$-$30$^\circ$11223 are young, which is consistent with the non-detection of objects with such high RV-shifts in the course of the MUCHFUSS project so far (Geier et al. \cite{geier11a,geier11b}). Most targets of this survey are faint subdwarfs located in the old halo population. The sdB-donor double-detonation channel is therefore predicted to occur in young stellar populations and contribute to the SN\,Ia population with short delay time (Ruiter et al. \cite{ruiter09}).

Given that systems like CD$-$30$^\circ$11223 are progenitors of some thermonuclear SN, a rough estimate can be made about the rate of such events. CD$-$30$^\circ$11223 is one out of $\sim100$ solved sdB binaries (Geier et al. \cite{geier11a}). About $50\%$ of the known sdB stars are in close binary systems. So we can estimate the number fraction of systems like CD$-$30$^\circ$11223 to be about $0.5\%$ of the whole population of sdB stars. According to binary evolution calculations, the birthrate of such stars in our Galaxy is $\sim5\times10^{-2}\,{\rm yr^{-1}}$ (Han et al. \cite{han03}). We therefore estimate the number of progenitor systems and the resulting SN\,Ia rate to be $\sim2.5\times10^{-4}\,{\rm yr^{-1}}$. This is consistent with the theoretical birthrate predicted for the WD+He star channel ($\sim3\times10^{-4}\,{\rm yr^{-1}}$, Wang et al. \cite{wang10}) But more importantly, it is smaller than the SN\,Ia birthrate of $\sim3\times10^{-3}\,{\rm yr^{-1}}$ and therefore consistent with observations (Capellaro \& Turato \cite{capellaro97}).

Although sub-Chandrasekhar scenarios in general have no well defined explosion mass, the parameter space for the sdB binary progenitors turns out to be quite narrow. According to hydrodynamic simulations the minimum mass of the WD should be $\sim0.8\,M_{\rm \odot}$, because carbon burning is not triggered for objects of much lower mass (Sim et al. \cite{sim12}). On the other hand, the WD must consist of carbon and oxygen to be able to explode as SN\,Ia. This limits the mass to values lower than $\sim1.1\,M_{\rm \odot}$, because even more massive WDs consist of oxygen, neon and magnesium and would rather collapse than explode. This mass range is further narrowed down by binary evolution calculations. Very close sdB+WD systems with companion masses around $0.8\,M_{\rm \odot}$ are predicted to be formed in much higher numbers than binaries with more massive companions. Another important constraint is that the timespan from the binary formation after the CE to the SN\,Ia explosion must be shorter than the core helium-burning lifetime ($\sim100\,{\rm Myr}$). Otherwise the sdB will turn into a WD before helium can be transferred. This restricts the orbital periods of possible sdB+WD progenitors to less than $\sim0.07\,{\rm d}$. 

The double-detonation scenario with hot subdwarf donor is the only proposed SN\,Ia scenario where both progenitors and remnants have been identified. Analysing a larger sample of those objects will allow us to put tight constraints on their properties and evolution.

\begin{acknowledgements}

Based on observations obtained at the European Southern Observatory, La Silla for programme 089.D-0265(A).\\
Based on observations with the William Herschel Telescope operated by the Isaac Newton Group at the Observatorio del Roque de los Muchachos of the Instituto de Astrofisica de Canarias on the island of La Palma, Spain.\\
Based on observations made at the South African Astronomical Observatory (SAAO).\\
Based on observations with the Southern Astrophysical Research (SOAR) telescope operated by the U.S. National Optical Astronomy Observatory (NOAO), the Ministério da Ciencia e Tecnologia of the Federal Republic of Brazil (MCT), the University of North Carolina at Chapel Hill (UNC), and Michigan State University (MSU).\\
We acknowledge the Director of SOAR for making the time for these observations during Technical and Engineering nights on the telescope.\\
A.I. acknowledges support from a research scholarship by the Elite Network of Bavaria.\\
V.S. acknowledges funding by the Deutsches Zentrum f\"ur Luft- und Raumfahrt (grant 50 OR 1110) and by the Erika-Giehrl-Stiftung.\\
S.G. and E.Z. are supported by the Deutsche Forschungsgemeinschaft (DFG) through grants HE1356/49-1 and HE1356/45-2, respectively.\\
Finally, we want to thank the anonymous referee for helpful comments and suggestions.

\end{acknowledgements}


\begin{thebibliography}{}

\bibitem[2011]{bloemen11} Bloemen, S., Marsh, T. R., \O stensen, R. H., et al. 2011, MNRAS, 410, 1787
\bibitem[2005]{brown05} Brown, W. R., Geller, M. J., Kenyon, S. J., \& Kurtz M. J. 2005, ApJ, 622, L33
\bibitem[1997]{capellaro97} Capellaro, E. \& Turatto, M. 1997, NATO ASI Series, 486, 77
\bibitem[2011]{claret11} Claret, A., \& Bloemen, S. 2011, A\&A, 529, 75
\bibitem[2010]{copperwheat10} Copperwheat, C., Marsh, T. R., Dhillon, V. S., et al. 2010, MNRAS, 402, 1824
\bibitem[2005]{edelmann05} Edelmann, H., Napiwotzki, R., Heber, U., Christlieb, N. \& Reimers, D. 2005, ApJ, 634, L181
\bibitem[1971]{eggleton71} Eggleton, P. P. 1971, MNRAS, 151, 351
\bibitem[1972]{eggleton72} Eggleton, P. P. 1972, MNRAS, 156, 361
\bibitem[1973]{eggleton73} Eggleton, P. P. 1973, MNRAS, 163, 279
\bibitem[2010]{fink10} Fink, M., R\"opke, F. K., Hillebrandt, W., et al. 2010, A\&A, 514, 53
\bibitem[2001]{fontaine01} Fontaine, G., Brassard, P., \& Bergeron, P. 2001, PASP, 113, 409
\bibitem[2007]{geier07} Geier, S., Nesslinger, S., Heber, U., et al. 2007, A\&A, 464, 299
\bibitem[2010]{geier10} Geier, S., Heber, U., Podsiadlowski, Ph., et al. 2010, A\&A, 519, 25
\bibitem[2011]{geier11a} Geier, S., Hirsch, H., Tillich, A., et al. 2011a, A\&A, 530, 28
\bibitem[2011]{geier11b} Geier, S., Maxted, P. F. L., Napiwotzki, R., et al. 2011b, A\&A, 526, 39
\bibitem[2012]{geier12} Geier, S., Marsh, T. R., Dunlap, B. H., et al. 2012, ASP Conf. Ser., in press (arXiv:1209.4740)
\bibitem[1994]{han94} Han, Z., Podsiadlowski, Ph., \& Eggleton, P. P. 1994, MNRAS, 270, 121
\bibitem[2000]{han00} Han, Z., Tout, C. A., \& Eggleton, P. P. 2000, MNRAS, 319, 215
\bibitem[2002]{han02} Han, Z., Podsiadlowski, Ph., Maxted, P. F. L., Marsh, T. R., \& Ivanova, N. 2002, MNRAS, 336, 449
\bibitem[2003]{han03} Han, Z., Podsiadlowski, Ph., Maxted, P. F. L., \& Marsh, T. R. 2003, MNRAS, 341, 669
\bibitem[2000]{heber00} Heber, U., Reid, N., \& Werner, K. 2000, MNRAS, 363, 198
\bibitem[2013]{heber13} Heber, U., Geier, S., G\"ansicke, B. 2013, EPJ Web of Conferences, 43, 04002 (arXiv:1211.5315)
\bibitem[2005]{hirsch05} Hirsch, H. A., Heber, U., O'Toole, S. J., \& Bresolin, F. 2005, A\&A, 444, L61
\bibitem[2013]{irrgang13} Irrgang, A., Wilcox, B., Tucker, E., \& Schiefelbein, L. 2013, A\&A, 549, 137
\bibitem[2009]{justham09} Justham, S., Wolf, C., Podsiadlowski, Ph., \& Han, Z. 2009, A\&A, 493, 1081
\bibitem[2012]{kilic12} Kilic, M., Brown, W. R., Allende Prieto, C., et al. 2012, ApJ, 751, 141
\bibitem[2009]{kovetz09} Kovetz, A., Yaron, O., \& Prialnik, D. 2009, MNRAS, 395, 1857
\bibitem[2010]{kromer10} Kromer, M., Sim, S. A., Fink, M., et al. 2010, ApJ, 719, 1067
\bibitem[1971]{landau71} Landau, L. D., \& Lifshitz, E. M. 1971, Classical theory of fields  (Pergamon Press, Oxford)
\bibitem[2000]{maxted00} Maxted, P.F.L., Marsh, T.R., \& North, R.C. 2000, MNRAS, 317, L41
\bibitem[2001]{maxted01} Maxted, P. F. L., Heber, U., Marsh, T. R., \& North, R. C. 2001, MNRAS, 326, 1391
\bibitem[2004a]{napiwotzki04a} Napiwotzki, R., Karl, C. A., Lisker, T., et al. 2004a, Ap\&SS, 291, 321
\bibitem[2004b]{napiwotzki04b} Napiwotzki, R., Yungelson, L., Nelemans, G., et al. 2004b, ASP Conf. Ser., 318, 402
\bibitem[2009]{nelemans09} Nelemans, G. 2009, Class. Quantum Grav., 26, 094030
\bibitem[2012]{nemeth12} N\'emeth, P., Kawka, A., \& Vennes, S. 2012, MNRAS, 427, 2180
\bibitem[1982]{nomoto82} Nomoto, K. 1982, ApJ, 257, 780
\bibitem[2006]{pollacco06} Pollacco, D. L., Skillen, I., Collier Cameron, A., et al. 2006, PASP, 118, 1407
\bibitem[1995]{pols95} Pols, O. R., Tout, C. A., Eggleton, P. P., \& Han, Z. 1995, MNRAS, 274, 964
\bibitem[1998]{pols98} Pols, O. R., Schr\"oder, K.-P., Hurley, J. R., Tout, C. A., \& Eggleton, P. P. 1998, MNRAS, 298, 525
\bibitem[2007]{roelofs07} Roelofs, G. H. A., Groot, P. J., Benedict, G. F., et al. 2007, ApJ, 666, 1174
\bibitem[2010]{roeser10} Roeser, S., Demleitner, M., \& Schilbach, E. 2010, AJ, 139, 2440
\bibitem[2009]{ruiter09} Ruiter, A. J., Belczynski, K., \& Fryer, C. 2009, ApJ, 699, 2026
\bibitem[1987]{shakura87} Shakura, N., I., \& Postnov, K. A. 1987, A\&A, 183, L21 
\bibitem[2010]{sim10} Sim, S., R\"opke, F. K., Hillebrandt, W., et al. 2010, ApJ, 714, L52
\bibitem[2012]{sim12} Sim, S., Fink, M., Kromer, M., et al. 2012, MNRAS, 420, 3003
\bibitem[2011]{vennes11} Vennes, S., Kawka, A., \& N\'emeth, P. 2011, MNRAS, 410, 2095
\bibitem[2012]{vennes12} Vennes, S., Kawka, A., O'Toole, S. J., N\'emeth, P, \& Burton, D. 2012, ApJ, 759, L25 (arXiv:1210.1512)
\bibitem[1988]{verbunt88} Verbunt, F., \& Rappaport, S. 1988, ApJ, 332, 193
\bibitem[2009a]{wang09a} Wang, B., Chen, X., Meng, X., \& Han, Z. 2009a, ApJ, 701, 1540
\bibitem[2009]{wang09c} Wang, B., \& Han, Z. 2009, A\&A, 508, L27
\bibitem[2009b]{wang09b} Wang, B., Meng, X., Chen, X., \& Han, Z. 2009b, MNRAS, 395, 847
\bibitem[2010]{wang10} Wang, B., Liu, Z., Han, Y., et al. 2010, Sci. China Ser. G, 53, 586
\bibitem[2012]{wang12} Wang, B., \& Han, Z. 2012, New Astronomy Reviews, 56, 122
\bibitem[2003]{winkler03} Winkler, F. P., Gupta, G., \& Long, K. S. 2003, ApJ, 585, 324
\bibitem[1986]{woosley86} Woosley, S. E., Taam, R. E., \& Weaver, T. A. 1986, ApJ, 301, 601
\bibitem[2003]{yoon03} Yoon, S.-C., \& Langer, N. 2003, A\&A, 412, 53
\end{thebibliography}
\end{document}